 \documentclass[final,5p,times,twocolumn]{elsarticle}

\usepackage{amssymb}
\usepackage{amsthm}
\usepackage{hyperref}
\usepackage{graphicx}
\usepackage{xcolor}
\usepackage{lineno}
\journal{Nuclear Instruments \& Methods in Physics Research, Section A.}
\begin{document}
\baselineskip 0pc



\title{Simulation studies related to the particle identification by the forward and backward RICH detectors at Electron Ion Collider}

\begin{frontmatter}
\cortext[cor1]{Corresponding author}
\author[INFNTrs]{D.S.Bhattacharya}
\author[INFNRoma1]{E.Cisbani}
\author[INFNTrs]{C. Chatterjee\corref{cor1}}
\ead{chandradoy.chatterjee@ts.infn.it}
\author[INFNTrs]{S. Dalla Torre}
\author[Duke]{C. Dilks}
\author[BNL]{A. Kiselev}
\author[SBrook]{H.Klest}
\author[INFNBol]{R. Preghenella}
\author[Duke]{A. Vossen}
\affiliation[INFNTrs]{organization={INFN Trieste}, 
            country={Sezione di Trieste, I-34149 Trieste, Italy}
            }
\affiliation[Duke]{organization={Duke University},
            country={ Durham, North Carolina 27708, U.S.A}
            }
\affiliation[INFNBol]{organization={INFN Bologna},
            country={Sezione di Bologna, I-40127 Bologna, Italy}
            }
\affiliation[INFNRoma1]{organization={INFN Roma 1},
            country={Sezione di Roma, I-00185 Roma, Italy}
            }
\affiliation[BNL]{organization={Brookheaven National Laboratory},
            country={Upton, New York 11973, U.S.A}
            }
\affiliation[SBrook]{organization={StonyBrook University},
            country={ Stony Brook, New York 11794, U.S.A.}
            }

\begin{abstract}
The Electron-Ion collider (EIC) will be the ultimate facility to study the dynamics played by colored quarks and gluons in the phenomenology of nucleons and nuclei, described by Quantum Chromodynamics. The physics programs will greatly rely on efficient particle identification (PID) in both the forward and the backward regions. The forward and the backward RICHes of the EIC have to be able to cover wide acceptance and momentum ranges; in the forward region a dual radiator RICH (dRICH) is foreseen and in the backward region a proximity-focusing RICH can be foreseen to be employed. The geometry and the performance studies of the dRICH have been performed as prescribed in the EIC Yellow Report using the ATHENA software framework. This part of our work reports the effort following the call for EIC detector proposal and the studies related to the forward and the backward RICH performance. 
In the forward region,the dRICH performance showed a pion- kaon separation from around 1 GeV/c to 50 GeV/c at a three sigma level;  the proximity focusing RICH (pfRICH) foreseen for the backward region can reach three sigma separation up to 3 GeV/c for e/$\pi$ and up to 10 GeV/c for $\pi$/K mass hypotheses.
\end{abstract}
\end{frontmatter}





\section{Requirements of particle identification in EIC}
\par
The EIC at the Brookhaven National Laboratory(USA)\cite{EICYR} is expected to start data-taking in the early 2030s to answer critical questions related to Quantum ChromoDynamics (QCD). A state-of-the-art machine and cutting-edge detector technology are fundamental for its success. Highly polarized electrons ($\sim$70-80\%) will be collided with highly polarized nucleons and light nuclei ($\sim$70-80\%) and also with heavy ions. The high luminosity collisions ($\sim$10$^{34}$  cm$^{-2}$s$^{-1}$ electron-proton) will take place over a wide center of mass energy (20-141 GeV) with a possibility of more than one interaction point. 
Much of EIC physics requires excellent particle identification over a wide phase space. The requirements are documented as a common effort of the entire EIC community; afterwards, we will refer to this report as the Yellow Report (YR) \cite{EICYR}. In response to the call for detector proposal submission, three proto-collaborations ATHENA \cite{athena}, ECCE \cite{ecce} and CORE\cite{CORE} had been formed. Our report mainly focuses on the simulation studies made using the ATHENA software framework for studying the forward and backward RICH detectors' performances.

\section{PID performance requirements for EIC physics}
The YR gives indicative requirements for the hadron PID in the electron-going endcap: better than 3$\sigma$ $\pi$/K up to 10 GeV/c \cite{EICYR}. 
The forward dRICH is aimed to perform 3$\sigma$ $\pi$/K separation up to $50$ GeV/c and e/$\pi$ separation up to 15 GeV/c. As mentioned in the YR, the required acceptance for the dRICH is $1.0\leq\eta\leq3.5$. 
\begin{figure}[!thb]
    \includegraphics[trim={1cm 0.5cm 1cm 1cm},clip,width=0.5\textwidth]{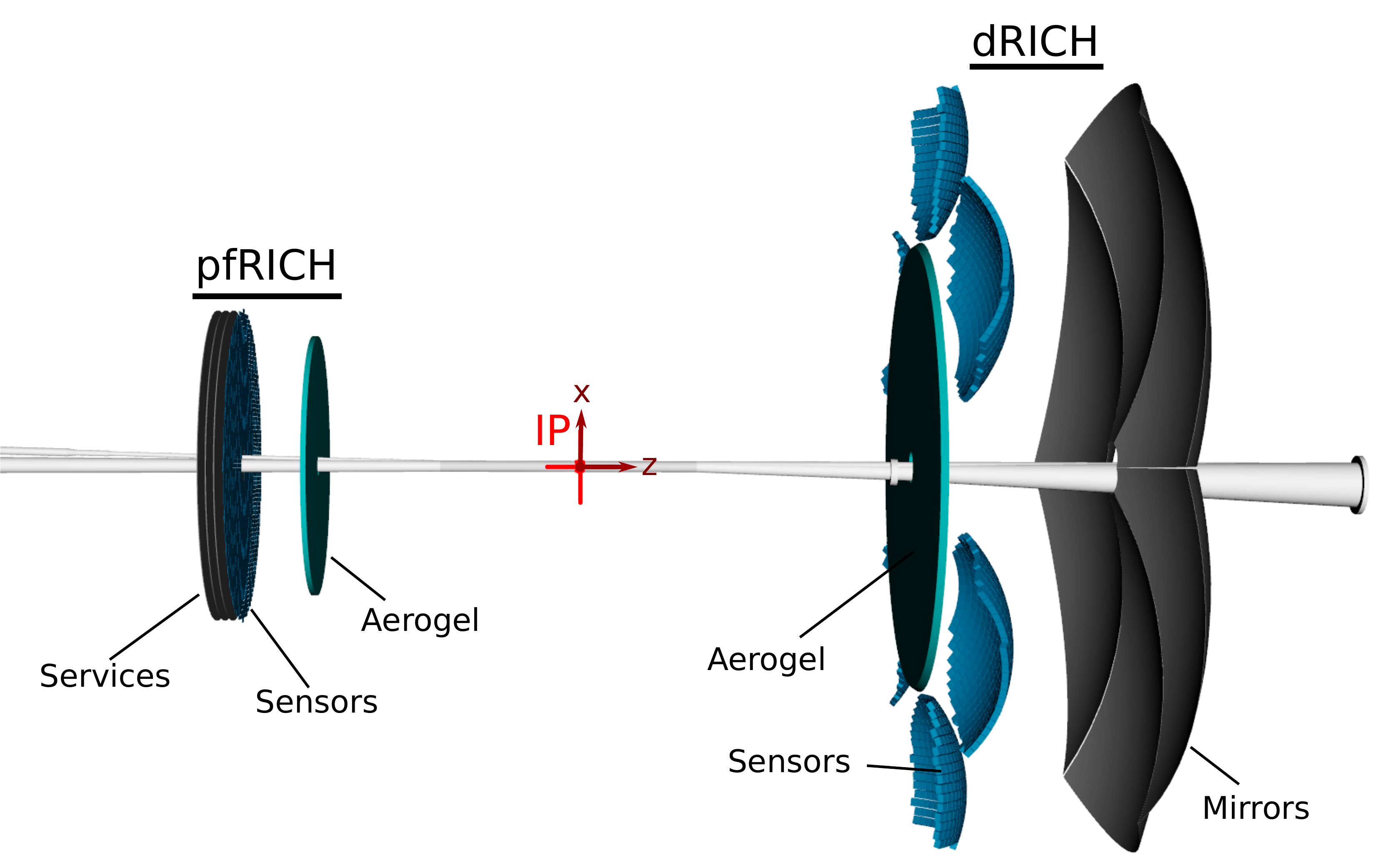}
    \vskip -1.0 em
    \caption{Schematic of the forward dual radiator RICH (dRICH) and backward proximity focusing (pfRICH) with respect to the interaction point (IP).}
    \label{fig:photoRICHes}
\end{figure}
Fig. \ref{fig:photoRICHes} shows the location of the two RICH detectors with respect to the interaction point (IP).
\section{pfRICH geometry}
The ATHENA design assumed that the proximity-focusing RICH will occupy space from -150~cm to -210~cm from the nominal IP (available space between the central tracker and crystal calorimeter). In our implementation, the vessel has a cylindrical shape, with an outer radius of 93~cm, and a cutaway at small radii as determined by the design of the beam pipe. 3~cm thick aerogel with an average refractive index $\langle n\rangle = 1.019$ was used in the simulations as the main Cherenkov radiator. The properties of the aerogel (refractive index variation, absorption, and Rayleigh scattering parameterizations as a function of wavelength) were taken from the available CLAS12 data \cite{CLAS12}. A 40~cm long expansion volume was assumed to be filled with $C_4F_{10}$ to provide an additional e/$\pi$ separation capability in a threshold mode below $\sim$ 2.9 GeV/c, with parameterizations taken from \cite{C4F10_ref1}. Hamamatsu S13361-3050AE-08 8x8 SiPM panels \cite{SiPM} are anticipated as a reference photosensor (3~mm single SiPM size). The SiPM Photon Detection Efficiency (PDE) as well as the geometric fill factor is taken according to the Hamamatsu specifications \cite{SiPM-specs}. About 15~cm of space behind the SiPM plane is reserved inside the vessel for the readout electronics and services. We applied an additional safety factor of 0.7, namely, we assume that only 70\% of photons that pass the PDE and the geometric fill-factor selection are actually detected and used in the Cherenkov angle evaluation.
\section{Forward dual radiator RICH (dRICH)}
In general, the dRICH configuration is very similar to the one from the YR. A substantial effort was made to accommodate such an apparatus in the overall tight space available for the ATHENA detector in the RHIC IP6 Interaction Region and to quantify its expected performance.
\subsection{dRICH location and vessel boundaries}
\begin{figure}[!thb]
    \centering
    \includegraphics[trim={0.3cm 8.75cm 16cm 0.1cm},clip,width=0.5\textwidth]{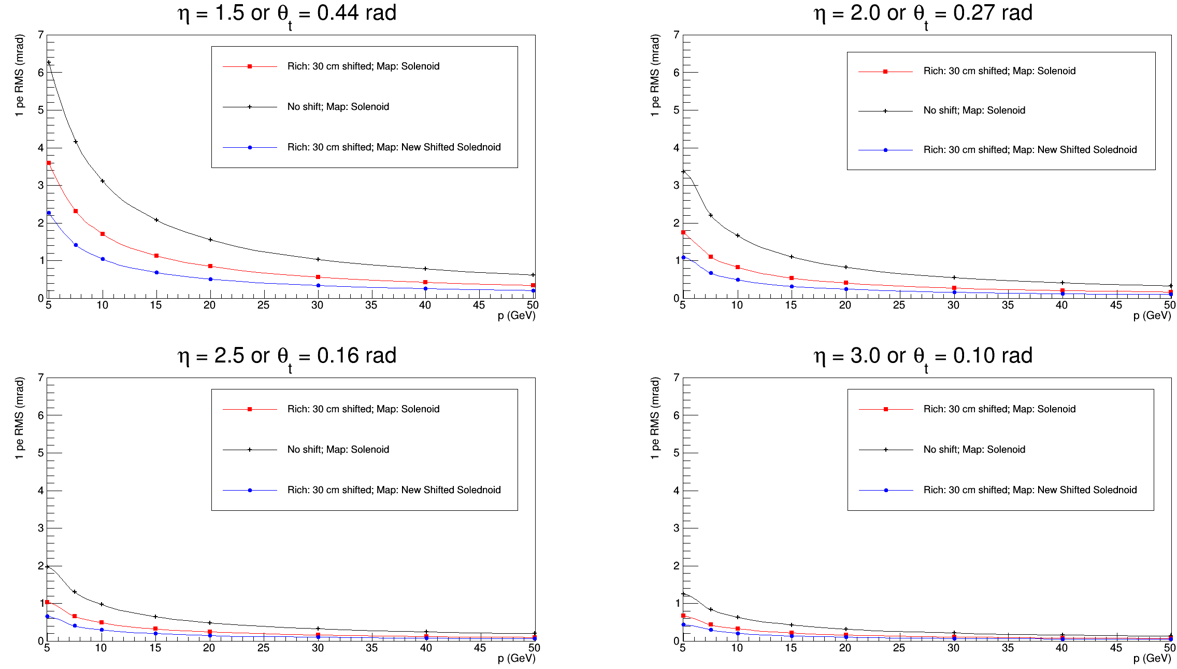}
    \vskip -1.5 em
    \caption{Example of the effect of the magnetic field on the single photon Cherenkov angle RMS. The black points represent dRICH located at nominal positions prior to finalization, the red points represent the dRICH additionally shifted by 30 cm away from the IP, and blue points represent the same 30 cm shift but with the solenoid coil additionally shifted by 25 cm towards electron endcap.}
    \label{fig:solenoid}
\end{figure}

During the optics tuning it was realized that the originally allocated space of $\sim$ 120~cm along the beamline is not sufficient to contain the focal plane inside the vessel, due to a very large polar angular acceptance. Therefore, it was necessary to shift the dRICH vessel further away from the IP in order to minimize the adverse effects of the solenoid fringe field, even though the coil and the return flux configuration of the new magnet were carefully tuned to observe the so-called projectivity requirement, namely to minimize the overall bending of the charged secondary particles originated from the IP. As a consequence of these studies, in the final ATHENA configuration, which is presented in the ATHENA proposal, the dRICH vessel was shifted by $\sim$30~cm away from the IP, and at the same time the solenoid coils were moved by 25~cm towards the electron-going endcap (see Fig. \ref{fig:solenoid}).  

\subsection{dRICH geometry}
The dRICH used the same aerogel type and the same SiPM sensors as the pfRICH. The safety factor of 70\% was also included. The vessel length along the beam line is 140~cm, and it occupies the range from +190~cm to +330~cm from the IP. The detector will have six $60^\circ$ sectors, each equipped with its own spherical mirror segment and a sensor plane. The aerogel thickness is 4~cm. We have used C$_2$F$_6$ gas as a radiator. This allows achieving a comfortable momentum overlap with the aerogel, and a 3$\sigma$ $\pi$/K separation is obtained up to $\sim$50~GeV/c. A C$_2$F$_6$ refractive index parameterization is taken from \cite{C2F6_ref1, C2F6_ref2}, and a conservative value of 10~m absorption length is taken for the simulations. The sensors are positioned on a sphere, with a square tiling algorithm. Three variables parameterize the spherical mirrors: the $z$ position of the backplane, which is the maximum $z$ the spherical mirror will reach, along with two focus tune parameters $f_x$ and $f_z$. Point-to-point focusing of the IP on the center of the sensor sphere corresponds to focus tunes $f_x=f_z=0$, which represents a starting point for the focus tuning; because the IP is far from the optical axis of the spherical mirror, spherical aberrations cause the proper point-to-point focal region to be significantly blurred. In order to focus Cherenkov rings on the sensors, parallel-to-point focusing is used. By changing the values of $f_x$ and $f_z$ it is possible to steer the parallel-to-point focal region to be as close to the sensor surfaces as possible. 
\section{Simulation studies for ATHENA}
A DD4Hep-based \cite{dd4hep} framework was used for the ATHENA simulation. The reconstruction was based on the Juggler framework \cite{Juggler}. A newly written Inverse Ray Tracing code (IRT), equally applicable in a standalone GEANT4 environment and in the ATHENA software framework. The code represents a substantial generalization of the IRT algorithm initially developed for the HERMES dual radiator RICH \cite{HERMES}. It allows one to perform ray tracing between the detected photon location in 3D space and the expected emission point range along the charged particle trajectory on a predefined sequence of refractive and reflective boundaries, using a 2D iterative Newton-Gauss minimization procedure. It is fully configurable and has a persistent model, which allows one to export and import ROOT files with the actual optics description. The code was originally developed to quantify the performance of the LHCb-like RICH-1 \cite{LHCb} configuration for the ATHENA forward RICH, with spherical and flat mirrors in a sequence, which cannot be easily handled by a simple 2D IRT algorithm. However, it is equally applicable to a simpler pfRICH geometry, where both absorption and Rayleigh scattering in the aerogel and refraction on the aerogel-gas boundary still play a role in the unbiased Cherenkov angle evaluation. The IRT library \cite{IRT}  is available in the ATHENA software repository, together with the standalone GEANT4 stepping code and the so-called Juggler plugin in the ATHENA reconstruction environment. IRT algorithm implementation in the ATHENA Juggler PID plugin is complemented by a sophisticated logic, performing sampling along the charged particle trajectory, which allows one to estimate the average Cherenkov angle in the same way for both straight tracks and in presence of a relatively strong bending component of the ATHENA solenoid magnetic field. A properly weighted ensemble of the Cherenkov photon angle estimates is then checked against e/$\pi$/K/p mass hypotheses, and a probability of each hypothesis is provided as an output.
\section{Performance Studies}
\subsection{pFRICH consistency}
\begin{figure}[!thb]
    \includegraphics[trim={0.5cm 0.1cm 0.85cm 0.25cm},clip,width=0.45\textwidth]{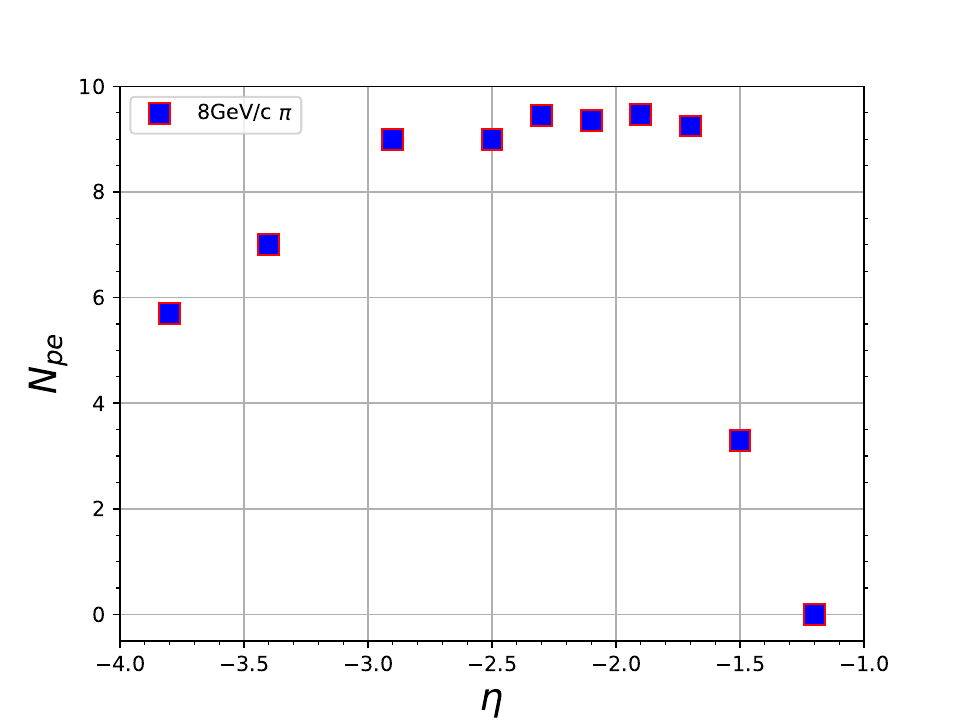}
    \vskip -1.0 em
    \caption{Number of detected photons in pfRICH rings (aerogel) as a function of pseudorapidity. The sharp drop at the edges is due to the containment of the partial ring in the sensors due to acceptance. 50\% of the ring contained in the sensor plane is defined as the acceptance limit of the detector. The numbers are estimated applying 70\% safety factor.}
    \label{fig:pfRICHAcc}
\end{figure}

\begin{figure}[!thb]
    \includegraphics[trim={0.5cm 0.1cm 0.85cm 0.25cm},clip,width=0.45\textwidth]{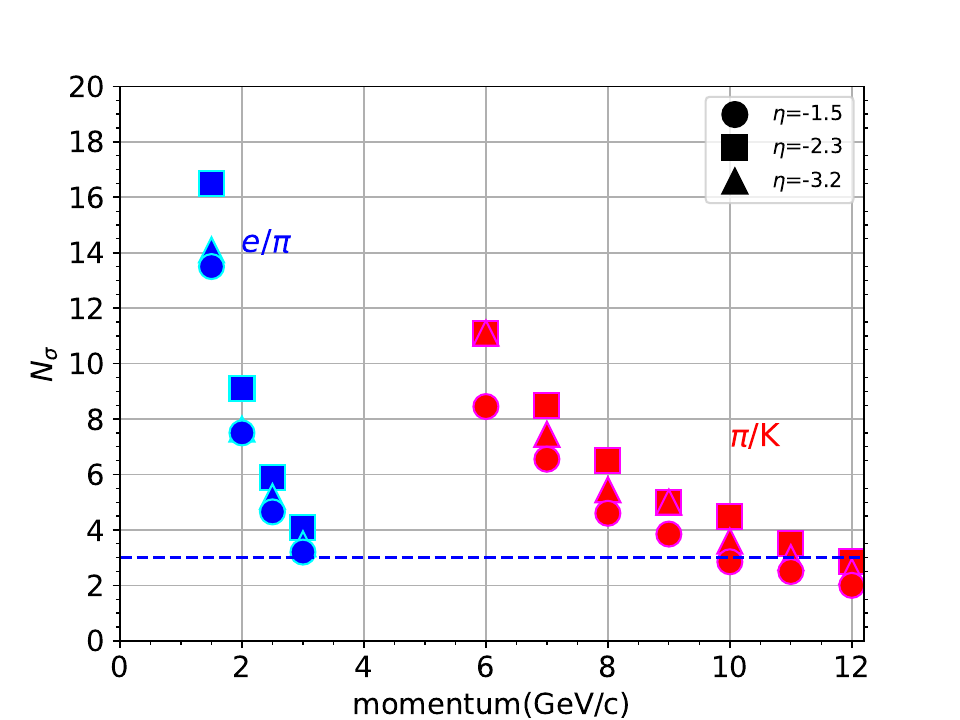}
    \vskip -1.0 em
    \caption{pfRICH N$_\sigma$ separation power as a function of momentum by using aerogel information.}
    \label{fig:pfRICHplots}
\end{figure}

To check the consistency of the software stack the performance studies of the pfRICH were done first, thanks to its simpler geometry. Reconstructed Cherenkov angles as a function of momentum have been studied for different mass hypotheses to check the consistency, and also for dRICH \cite{Suppl-Mat-Wiki}. Single particles (e/$\pi$/K) were simulated in turn to estimate the N$_\sigma$ separation as a function of momentum and pseudorapidity ($\eta$). It has been demonstrated that the Yellow Report requirements can be achieved using the simple pfRICH geometry and the reconstructed Cherenkov angles provide a satisfactory kaon rejection factor without diluting the pion identification efficiency. The acceptance plot shows that between $\eta$ 1.6 to 3 the number of detected photons is constant. We have defined 50\% of the maximum number of the detected photons as the working acceptance. This demonstrates the pfRICH can cover a region larger than $\eta$ of 3.5 in the backward region (see Fig. \ref{fig:pfRICHAcc}). For three different $\eta$ regions the N$_\sigma$ had been studied. One can reach three sigma separation up to 3 GeV/c for e/$\pi$ and up to 10 GeV/c for $\pi$/K mass hypothesis (see Fig. \ref{fig:pfRICHplots}). Using an equal mixture of pion and kaon samples for particles at saturation a pion rejection factor as a function of kaon detection efficiency is also computed. This shows that without diluting the kaon detection efficiency, a high level of pion rejection can be obtained (Fig. \ref{fig:dRICHEvaluation2} (left)).  
\subsection{dRICH performance}
\begin{figure}[!thb]
    \includegraphics[trim={1cm 0.5cm 1.3cm 1cm},clip,width=0.5\textwidth]{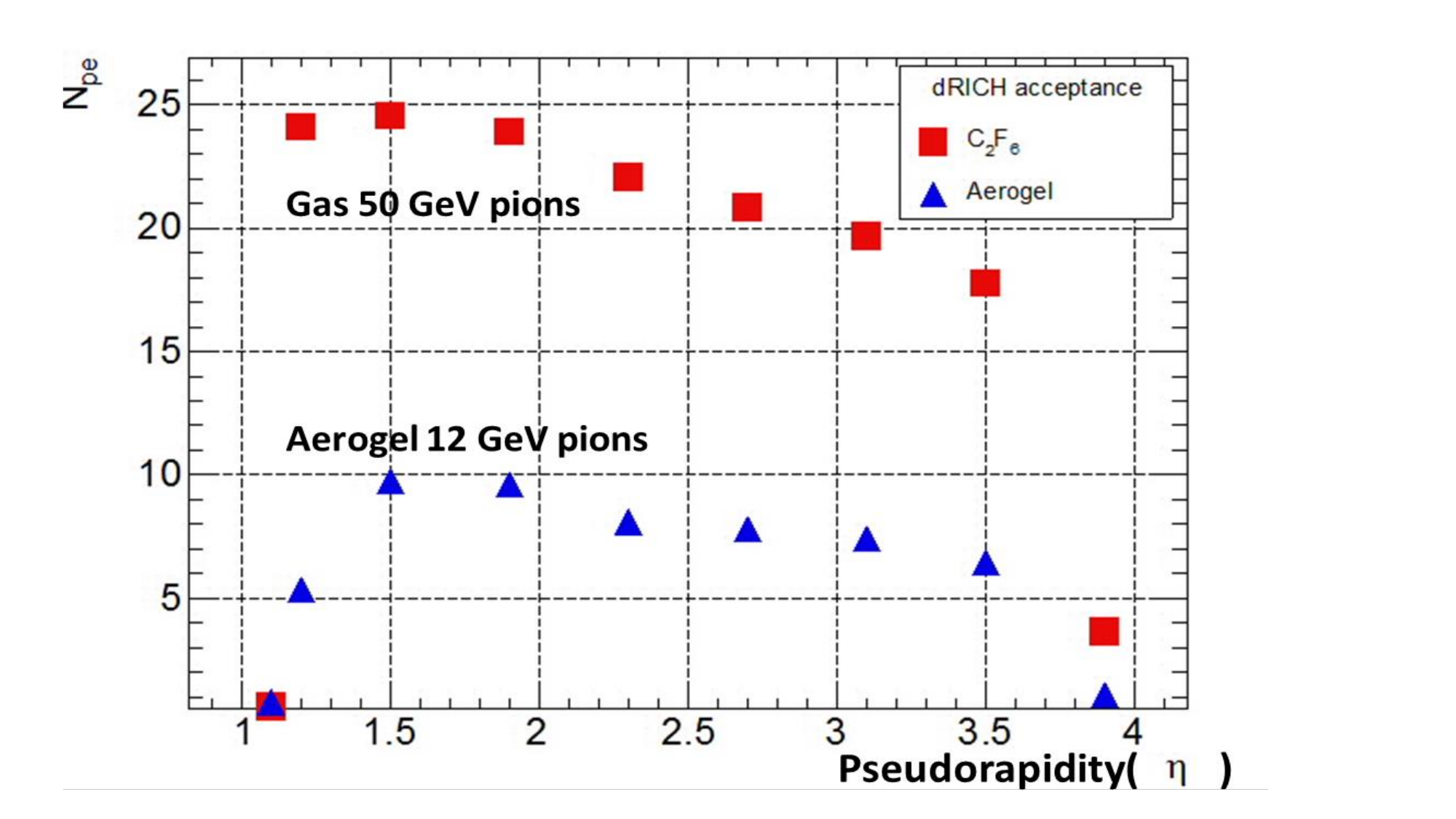}
    \vskip -1.0 em
    \caption{Number of detected photons in dRICH rings (aerogel and C$_2$F$_6$) as a function of pseudorapidity. The sharp drop at the edges is due to the containment of the partial ring in the sensors due to acceptance. 50\% of the ring contained in the sensor plane is defined as the acceptance limit of the detector.}
    \label{fig:dRICHAcc}
\end{figure}

\begin{figure}[!thb]
    \includegraphics[trim={1cm 0.5cm 0.6cm 1cm},clip,width=0.5\textwidth]{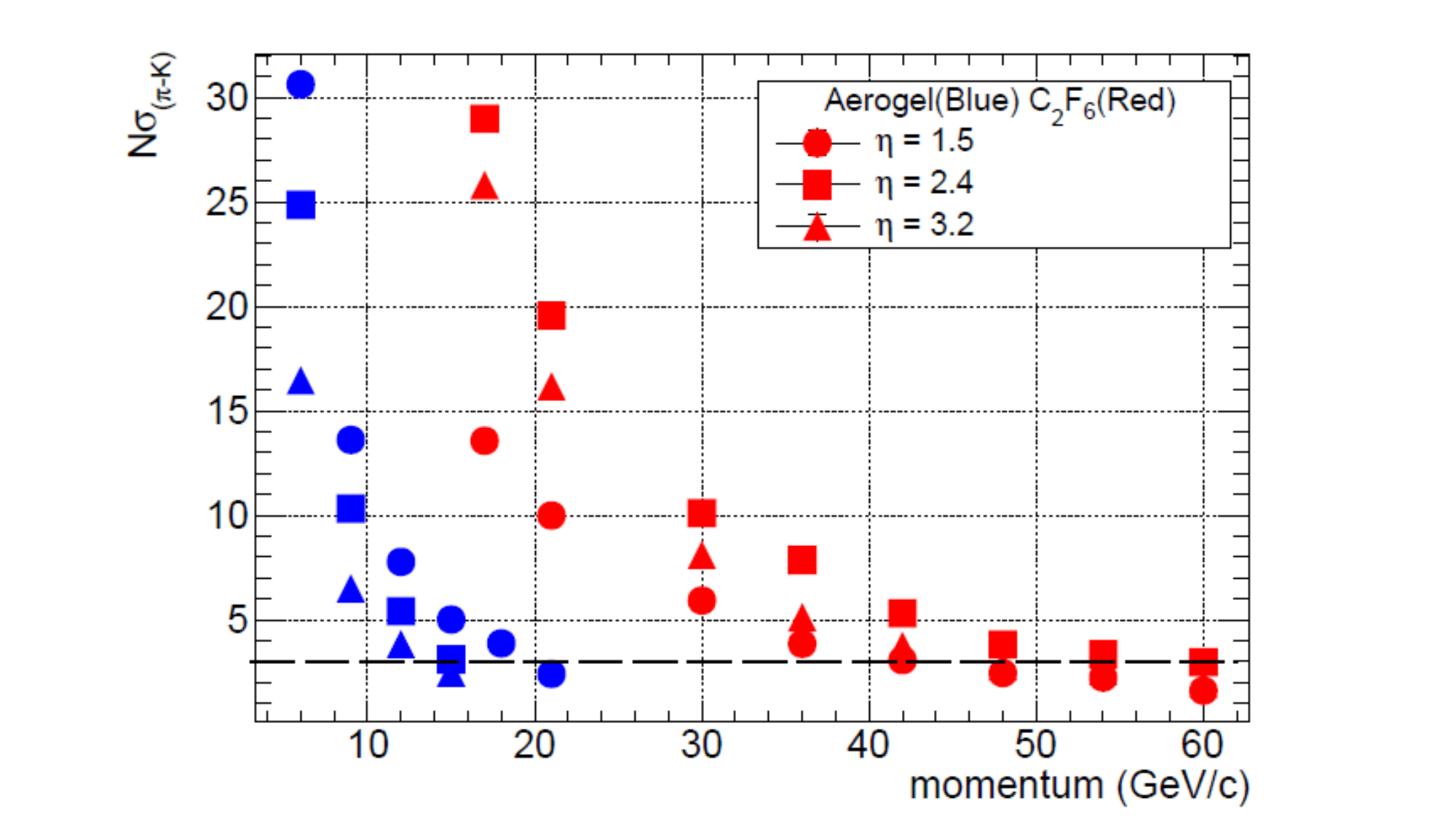}
    \vskip -1.0 em
    \caption{N$_\sigma$ separation as a function of momentum of positively identified $\pi$/K using aerogel and C$_2$F$_6$ information from dRICH.}
    \label{fig:dRICHplots}
\end{figure}

The forward dRICH also showed performance as prescribed by the YR. Both for the aerogel and gas we have observed the acceptance is from an $\eta$ 1.2 to around 3.5 (Fig. \ref{fig:dRICHAcc}). For the separation power, it is evident that using the aerogel and gas information once can achieve a continuous $\pi$/K separation from some hundreds of MeV/c up to 50 GeV/c without any loss in the performance (Fig. \ref{fig:dRICHplots}).
\begin{figure}[!thb]
    \includegraphics[trim={1cm 0cm 1.5cm 1cm},clip,width=0.55\textwidth]{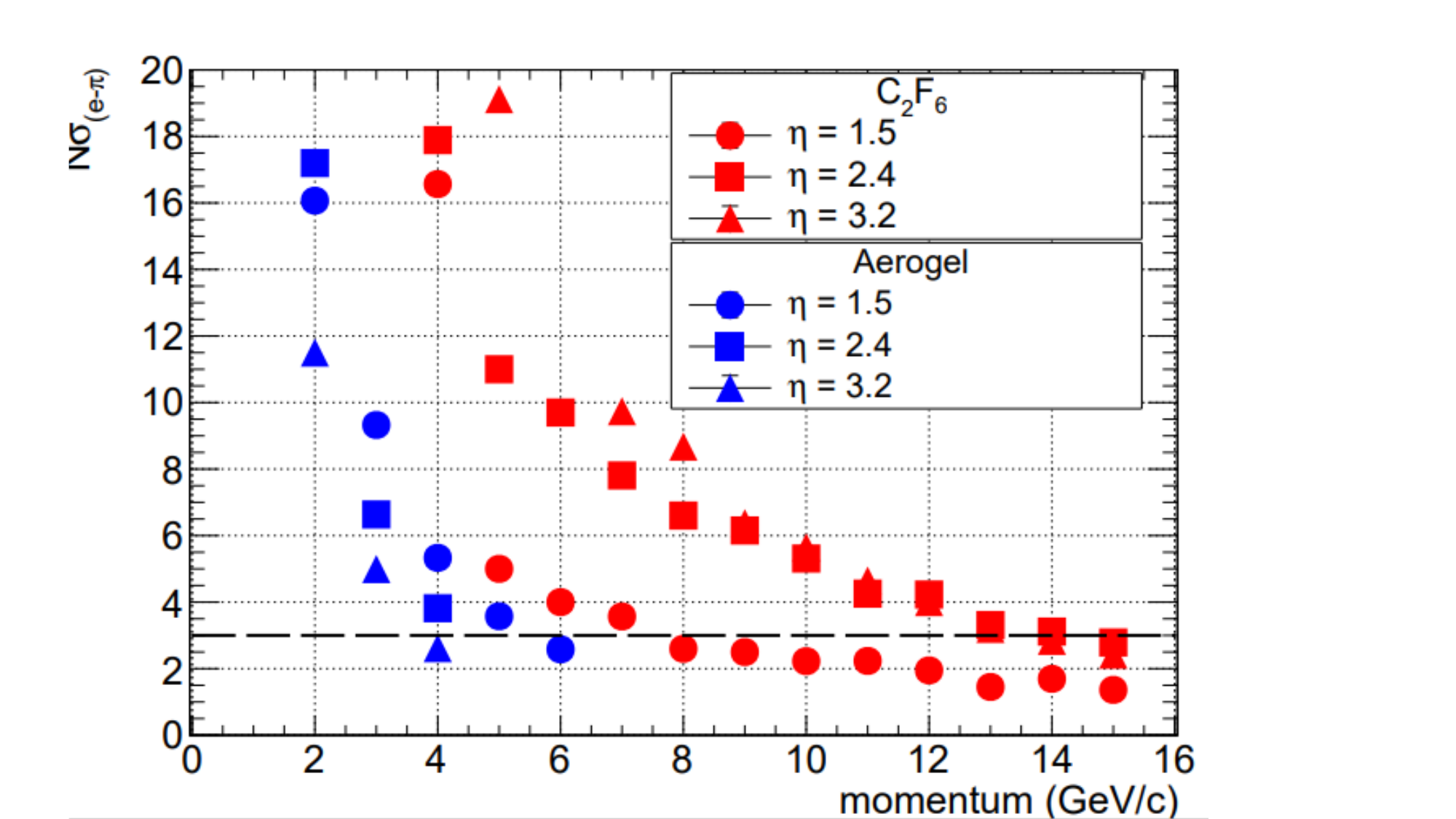}
    \vskip -0.5 em
    \caption{N$_\sigma$ separation of e/$\pi$ using aerogel and C$_2$F$_6$ information from dRICH.}
    \label{fig:dRICHplots2}
\end{figure}
\begin{figure}[!thb]
    \centering
    \includegraphics[trim={0cm 0.4cm 0cm 0cm},clip,width=0.45\textwidth]{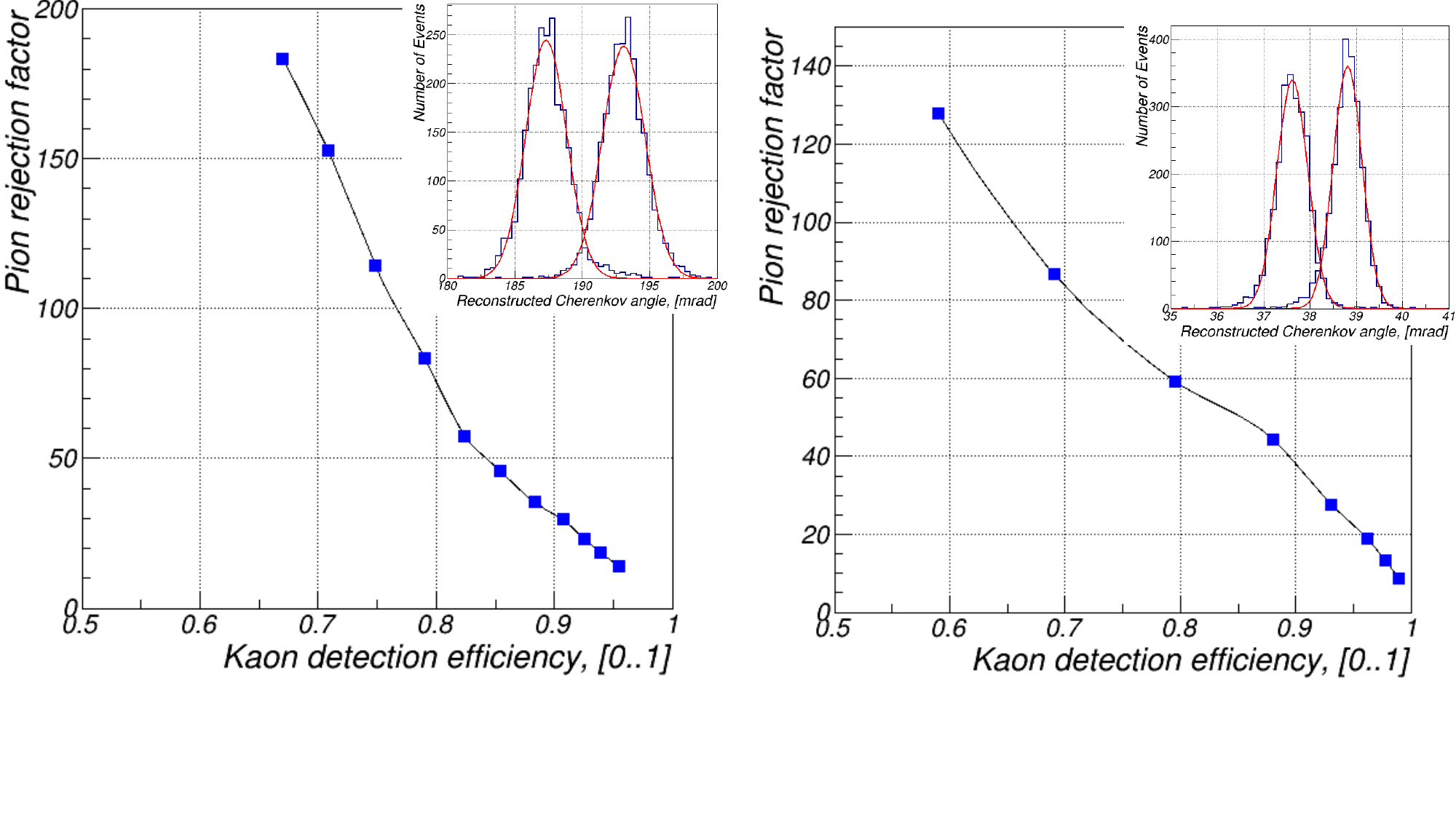}
    \vskip -3.0 em
    \caption{Pion rejection factor as a function of kaon identification efficiency for pfRICH (left) and dRICH (right). The corner plots in both panels show the reconstructed Cherenkov angles of a sample of an equal fraction of the pion/kaon mixture at saturated momentum and at a fixed pseudo-rapidity. The applied cut on the reconstructed angle is varied to estimate the kaon detection efficiency as a function of pion rejection factor.}
    \label{fig:dRICHEvaluation2}
\end{figure}

As already mentioned the forward dRICH should be able to reject electrons, and it has been demonstrated that up to 15~GeV/c the separation can be done from mid to high $\eta$ (Fig. \ref{fig:dRICHplots2}). Around $\eta$ =1.5 the separation power is diluted, as the effect of the solenoidal field plays a critical role in this region. Instead of using mirrors with a single radius of curvature, each sector can be equipped with a mirror divided into two sections with slightly different radii of curvature in order to obtain better resolution over the region. Stand-alone studies with dual mirrors have shown promising results. Similar to the pfRICH, the pion rejection factor as a function of kaon detection efficiency shows for saturated pions and kaons that dRICH is able to reject pions without any dilution of the kaon detection efficiency (see Fig. \ref{fig:dRICHEvaluation2} (right)). We have also obtained that the reconstructed mass of the particles as a function of particle momentum obtained from the reconstructed Cherenkov angle, reconstructed refractive index (using a pion mass hypothesis) and known particle momentum for both aerogel and C$_2$F$_6$ \cite{Suppl-Mat-Wiki}. 

\section{Conclusions and Future Plans}
In the ATHENA software framework, we have studied and demonstrated that the PID requirements can be achieved using two RICH detectors in the forward and backward endcaps by exploiting the emitted Cherenkov photons and their dependency on a  threshold momentum \cite{Suppl-Mat-Wiki}. The obtained N$_\sigma$ separation from simulation as a function of the particle momentum and pseudorapidity was translated into detection efficiencies to use in the Delphes \cite{delphes} framework for physics simulation. The reconstruction software can be more sophisticated in the future, and the implementation of likelihood-based PID algorithms and machine learning can be options for higher-level physics studies to perform PID. Nevertheless, the studies performed in the ATHENA framework are already promising. Independent of ATHENA software, the ECCE proto-collaboration has also demonstrated similar results using a modular RICH (mRICH) in the backward endcap and a dRICH with different geometrical parameters in the forward direction \cite{ecce}. Following recommendations of the detector committee advisory panel, the 1.5T magnet configuration has been chosen as the baseline configuration, it has been mentioned that both ECCE and ATHENA proto collaborations are capable of delivering the entire EIC physics program. Hence, the new EPIC collaboration has been formed taking advantage of lessons learnt by the both ECCE and ATHENA collaborations, within ePIC a re-optimization of the forward RICH is ongoing. For the tuning and characterization studies, single photon resolution, and the number of detected photons per particle will be used also for characterizing the RICH detectors of the ePIC collaboration \cite{Suppl-Mat-Wiki}.
\section{Acknowledgements}
One of the authors (C.Chattejee) is supported by the European Union’s Horizon 2020 Research and Innovation Programme under Grant Agreement AIDAinnova - No 101004761. The work of C. Dilks and A. Vossen is supported by the U.S. Department of Energy, Office of Science, Office of Nuclear Physics.





\end{document}